\documentclass[12pt]{article}

\usepackage{amssymb}
\usepackage{amsmath}

\usepackage{graphicx}

\begin{document} %
%%%%%%%%%%%%%%%%%%

\newcommand{\bs}{\boldsymbol} % bold symbols in mathematical mode

%%%%%%%%%%%%%%%%%%%%%%%%%%%%%%%%%%%%%%%%%%%%%%%%%%%%%%%%%%%%%%%%%%%%%

\title{Search for noncommutative interactions in
$\gamma\gamma\rightarrow\gamma\gamma$ process at the LHC}

\author{
S.C. \.{I}nan\thanks{Electronic address: sceminan@cumhuriyet.tr}
\\
{\small Department of Physics, Sivas Cumhuriyet University, 58140,
Sivas, Turkey}
\\
{\small and}
\\
A.V. Kisselev\thanks{Electronic address:
alexandre.kisselev@ihep.ru} \\
{\small Division of Theoretical Physics, A.A. Logunov Institute for
High Energy Physics,}
\\
{\small NRC ``Kurchatov Institute'', 142281, Protvino, Russia}}

\date{}

\maketitle

\begin{abstract}
The noncommutative QED (NCQED) has a non-Abelian nature due to the
presence of 3- and 4-photon vertices in the lagrangian. Thus, NCQED
predicts a new physics contribution to the
$\gamma\gamma\rightarrow\gamma\gamma$ scattering already at the tree
level. We have examined NCQED by studying the light-by-light process
at the 14 TeV LHC with intact protons. Our results show that the NC
scale up to $\Lambda_{\mathrm{NC}} = 1.65(1.35)$ TeV can be probed
in the $pp\rightarrow p(\gamma\gamma)p \rightarrow
p'(\gamma\gamma)p'$ collision for the time-space (space-space) NC
parameters. These bounds are stronger than the limits that can be
obtained in the light-by-light scattering at high energy linear
colliders.
\end{abstract}

%%%%%%%%%%%%%%%%%%%%%%%%%%%%%%%%%%%%%%%%%%%%%%%%%%%%%%%%%%%%%%%%%%%%%

%%%%%%%%%%%%%%%%%%%%%%%%
\section{Introduction} %
%%%%%%%%%%%%%%%%%%%%%%%%

The quantization of the electromagnetic field in a noncommutative
(NC) space-time has a long story \cite{Snyder:1947_1,Snyder:1947_2}.
To a large extent, an interest in NC quantum field theories
\cite{Landi:1997}-\cite{Szabo:2003} was mainly motivated by string
theory \cite{Connes:1994}-\cite{Akhmedov:2001}. In NC field theories
the conventional coordinates are represented by non-commutative
operators,
\begin{equation}\label{NC_space-time}
[\hat{X}_\mu, \hat{X}_\nu] = i\theta_{\mu\nu} \;,
\end{equation}
where $\theta_{\mu\nu}$ is the NC \emph{constant} of dimension
$(\mathrm{mass})^{-2}$. In what follows,
\begin{equation}\label{c_Lambda}
\theta_{\mu\nu} = \frac{c_{\mu\nu}}{\Lambda_{\mathrm{NC}}^2} \;,
\end{equation}
where the dimensionless elements of the antisymmetric matrix
$c_{\mu\nu}$ are assumed to be of order unity.

Let the field $\hat{\Phi}(\hat{X})$ be an element of the algebra
\eqref{NC_space-time}. The noncommutivity of the space-time can be
implemented by the \emph{Weyl-Moyal correspondence}
\cite{Weyl:1927}-\cite{Alvarez-Gaume:2001}
\begin{align}\label{W-M_correspondence}
\hat{\Phi}(\hat{X}) &= \frac{1}{(2\pi)^{2}} \int d^4x
\,e^{ik\hat{X}} \phi(k) \nonumber \\
\phi(k) &= \frac{1}{(2\pi)^{2}} \int d^4k \,e^{-ikx} \Phi(x) \;,
\end{align}
where $k$, $x$ are real variables. Thus, we associate
$\hat{\Phi}(\hat{X})$ with a function of classical variable $x$. As
it follows from \eqref{W-M_correspondence},
\begin{align}\label{W_M_product}
\hat{\Phi}_1(\hat{X})\hat{\Phi}_2(\hat{X}) &= \frac{1}{(2\pi)^{4}}
\int d^4k \,d^4p \,e^{ik\hat{X}} \phi(k) \,e^{ip\hat{X}} \phi(p) \;,
\nonumber \\
&= \frac{1}{(2\pi)^{4}} \int d^4k \,d^4p \,e^{i(k+p)\hat{X} - k^\mu
p^\nu [\hat{X}_\mu, \hat{X}_\nu]/2} \phi(k) \phi(p) \;,
\end{align}
where we used the Backer-Campbell-Hausdorff formula. Thus, the NC
version of a field theory is given by replacing field products by
the \emph{star product} defined as
\begin{align}\label{star_product}
\hat{\Phi}_1(\hat{X})\hat{\Phi}_2(\hat{X}) &\leftrightarrow
(\Phi_1 * \Phi_2)(x) \;, \nonumber \\
(\Phi_1 * \Phi_2)(x) &= \exp \!\!\left[\frac{i}{2}
\frac{\partial}{\partial \xi^\mu} \,\theta^{\mu\nu}
\frac{\partial}{\partial \eta^\nu} \right] \!\Phi_1(x + \xi)
\Phi_2(x + \eta) \bigg|_{\xi = \eta = 0} \;.
\end{align}
It obeys the associative law. To the leading order in $\theta$, the
star product is given by
\begin{equation}\label{leading order_theta}
\Phi_1 * \Phi_2 = \Phi_1 \Phi_2 + \frac{i}{2} \,\theta^{\mu\nu}
\partial_\mu \Phi_1 \partial_\nu \Phi_2 + \mathrm{O}(\theta^2) \;.
\end{equation}

It is useful to define a generalized commutator known as the
\emph{Moyal bracket} (MB) by the relation
\begin{equation}\label{Moyal_bracket}
[\Phi_1, \Phi_2]_{\mathrm{MB}} = \Phi_1*\Phi_2 - \Phi_2*\Phi_1 \;.
\end{equation}
As one can see, the MB of coordinates,
\begin{equation}\label{M_bracket_coordinates}
[x_\mu, x_\nu]_{\mathrm{MB}} = x_\mu * x_\nu - x_\nu * x_\mu \;,
\end{equation}
in agreement with the commutator relation on the NC space-time
\eqref{NC_space-time}.

There is a relation between the matrix $c_{\mu\nu}$ \eqref{c_Lambda}
and the Maxwell field strength, since in string theory the
quantization of NC quantum field theory is described by the
excitations of D-branes in the presence of the background EM field
\cite{Connes:1994}-\cite{Akhmedov:2001}. The $c_{0i}$ coefficients
are defined by the direction of a background electric field,
$\mathbf{E} = (c_{01}, c_{02}, c_{03})/\Lambda_{\mathrm{NC}}^2$. The
$c_{ij}$ elements are related to a background magnetic field,
$\mathbf{B} = (c_{23}, c_{02}, -c_{12})/\Lambda_{\mathrm{NC}}^2$.

Note that theories with nonzero $c_{0i}$ in \eqref{NC_space-time} do
not generally obey unitarity
\cite{Gomis:2000}-\cite{Chaichian:2001_1}. However, theories with
only space-space noncommutativity, $c_{ij} \neq 0$, $c_{0i} = 0$,
are unitary.

%%%%%%%%%%%%%%%%%%%%%%%%%%%%%%
\section{Noncommutative QED} %
%%%%%%%%%%%%%%%%%%%%%%%%%%%%%%

Noncommutative QED (NCQED), based on the group $U(1)$, has been
studied in a number of papers \cite{Martin:1999}-\cite{Morita:2002}.
It was shown that unbroken $U(N)$ gauge theory is both gauge
invariant and renormalizable at the one-loop level
\cite{Martin:1999}, \cite{Bonora:2001}. The pure noncommutative
$U(1)$ Yang-Mills action is defined as
\begin{equation}\label{action}
S_{\mathrm{NCQED}} = - \frac{1}{4e^2} \int d^4x F_{\mu\nu} *
F^{\mu\nu} \;,
\end{equation}
with
\begin{equation}\label{F_mu_nu}
F_{\mu\nu} = \partial_\mu A_\nu - \partial_\nu A_\mu - i [A_\mu,
A_\nu]_{\mathrm{MB}} \;.
\end{equation}
We see that even in the $U(1)$ case the potential $A_\mu$ couples to
itself. One can easily check that the action \eqref{action} is
invariant under $U(1)$ transformation defined as
\begin{equation}\label{U(1)_symmetry}
A_\mu(x) \rightarrow A'_\mu(x) = U(x)*A_\mu(x)*[U(x)]^{-1} + i
U(x)*\partial_\mu[U(x)]^{-1} \;,
\end{equation}
where
\begin{equation}\label{U(1)}
U(x) = e^{i\alpha(x)} = 1 + i\alpha(x) -
\frac{1}{2}\alpha(x)*\alpha(x) + \cdots \;.
\end{equation}
The covariant derivative
\begin{equation}\label{covariant_derivative}
D_\mu \varphi - \partial_\mu \varphi- i A_\mu *\varphi
\end{equation}
transforms covariantly. The $*$ product admits only the fields
$\varphi$ with charge 0 or $\pm 1$ \cite{Hayakawa:2000}. The field
strength transforms as
\begin{equation}\label{F_transformation}
F_{\mu\nu} \rightarrow F'_{\mu\nu} = U(x)*F_{\mu\nu}*[U(x)]^{-1} \;.
\end{equation}
Using relations $U*U^{-1} = U^{-1}*U = I$ and the cyclic property of
the star product under the integral \cite{Riad:2000}, we find that
\begin{equation}\label{action_final}
S_{\mathrm{NCQED}} = - \frac{1}{4e^2} \int d^4x F_{\mu\nu}
F^{\mu\nu} \;.
\end{equation}

\begin{figure}[htb]
\begin{center}
\includegraphics[scale=0.5]{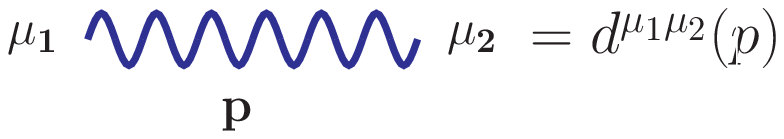} \\
\vspace{0.5cm}
\hspace{0.25cm}
\includegraphics[scale=0.5]{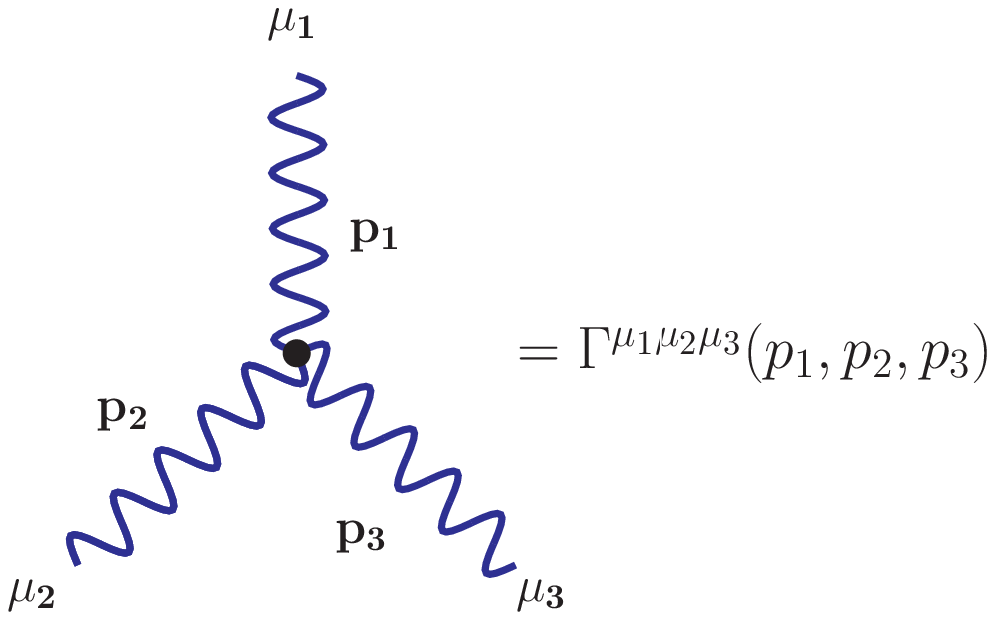} \\
\vspace{0.5cm} \hspace{1.5cm}
\includegraphics[scale=0.5]{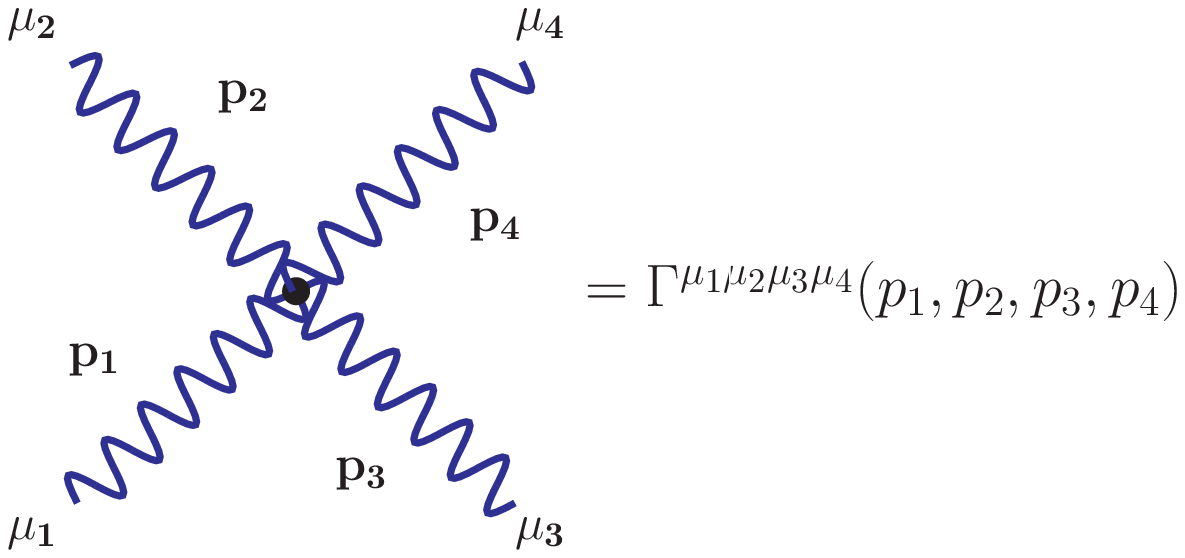}
\caption{The Feynman rules of the noncommutative QED.}
\label{fig:F_rules}
\end{center}
\end{figure}

The 2-point photon function is identical in NC and commutative
spaces, since the quadratic term in \eqref{action_final} remains the
same,
\begin{equation}\label{propagator}
d^{\mu_1\mu_2}(p) = - i\frac{g^{\mu_1\mu_2}}{p^2 + i\varepsilon} \;.
\end{equation}
Due to the presence of the $*$ product and MB bracket, the theory
reveals \emph{non-Abelian} nature. Namely, both 3-point and 4-point
photon vertices are generated. The Feynman rules of the pure NCQED
\cite{Armoni:2001}-\cite{Aref'eva:2001} are shown in
Fig.~\ref{fig:F_rules}
%
%%%%%%%%%%%%%%%%%%%%%%%%%%%%%%%%%%%%
% Figure 1. Feynman rules of NCQED %
%%%%%%%%%%%%%%%%%%%%%%%%%%%%%%%%%%%%
They are given by the following expressions
\begin{align}\label{3_point_vertex}
&\Gamma^{\mu_1\mu_2\mu_3}(p_1,p_2,p_3) = -2e\sin \left( \frac{1}{2}
\,p_1\wedge p_2\right) \nonumber \\
&\times \big[ (p_1 - p_2)^{\mu_3}g^{\mu_2\mu_3} + (p_2 -
p_3)^{\mu_1}g^{\mu_2\mu_3} + (p_3 - p_1)^{\mu_2}g^{\mu_3\mu_1} \big]
,
\end{align}
\begin{align}\label{4_point_vertex}
& \Gamma^{\mu_1\mu_2\mu_3\mu_4}(p_1,p_2,p_3,p_4)
\nonumber \\
&= -4ie^2 \bigg[ \left( g^{\mu_1\mu_3}g^{\mu_2\mu_4} -
g^{\mu_1\mu_4}g^{\mu_2\mu_3} \right) \sin \!\left( \frac{1}{2}
\,p_1\wedge p_2 \right) \sin \!\left( \frac{1}{2} \,p_3\wedge p_4
\right)
\nonumber \\
&\hspace{14mm} + \left( g^{\mu_1\mu_4}g^{\mu_2\mu_3} -
g^{\mu_1\mu_2}g^{\mu_3\mu_4} \right) \sin \!\left( \frac{1}{2}
\,p_1\wedge p_3 \right) \sin \!\left( \frac{1}{2} \,p_2\wedge p_4
\right)
\nonumber \\
&\hspace{14mm} + \left( g^{\mu_1\mu_2}g^{\mu_3\mu_4} -
g^{\mu_1\mu_3}g^{\mu_2\mu_4} \right) \sin \!\left( \frac{1}{2}
\,p_1\wedge p_4 \right) \sin \!\left( \frac{1}{2} \,p_2\wedge p_3
\right) \bigg] ,
\end{align}
where the wedge product is defined as
\begin{equation}\label{notation}
p\wedge k = p^\mu  k^\nu \theta_{\mu\nu} = \frac{p^\mu k^\nu
c_{\mu\nu}}{\Lambda_{\mathrm{NC}}^2} \;.
\end{equation}
Note that $p\wedge k = - k\wedge p$ and $p\wedge p = 0$. As one can
see, the Feynman rules are very similar to those in
\emph{non-abelian} gauge theory, with the structures constants
replaced by factors $2\sin[(p_i\wedge p_j)/2]$. These factors arise
as a consequence of the MB. Indeed, we have
\begin{equation}\label{MB_vs_NC_factor}
[A_\mu,A_\nu]_{\mathrm{MB}} = i\!\int \!d^4p_i \,d^4p_j \,e^{i(p_i +
p_j)x}A_\mu(p_i)A_\nu(p_j) \left[ 2\sin \!\left( \frac{1}{2}
\,p_i\wedge p_j \right)\right] .
\end{equation}
Thus, NCQED predicts new tree-level contributions to the
$\gamma\gamma\rightarrow\gamma\gamma$ collision.

Note that $c_{\mu\nu}$ \eqref{c_Lambda} in not a tensor. It means
that the Lorentz symmetry is explicitly violated. However, it is
quite different from Lorentz breaking models discussed often in the
literature, since it can hold only at energies of order
$\Lambda_{\mathrm{NC}}$. Moreover, since the NCQED is CPT invariant,
available experimental bounds on observables which are
\emph{simultaneously} CPT and Lorentz violating cannot be used to
constrain NCQED.

A number of noncommutative extensions of the Standard Model is
constructed \cite{Lizzi:1997}-\cite{Besnard:2021}. Since we are
interested in the light-by-light collision, we will work in the
framework of the NCQED.

%%%%%%%%%%%%%%%%%%%%%%%%%%%%%%%%%%%%%%%%%%%%%%%%
\section{Light-by-light scattering at the LHC} %
%%%%%%%%%%%%%%%%%%%%%%%%%%%%%%%%%%%%%%%%%%%%%%%%

The observable signatures in a number of NCQED processes in $e^+e^-$
collisions has been considered in
\cite{Mathews:2001}-\cite{Ohl:2004}. The light-by-light (LBL)
scattering in ultraperipheral Pb+Pb collisions in the NCQED context
have been recently studied in \cite{Horvat:2020,Latas:2021}. Our
goal is to examine the LBL scattering in $pp$ collisions at the 14
TeV LHC through the process $pp\rightarrow p(\gamma\gamma)p
\rightarrow p'(\gamma\gamma)p'$. Here the final state photons are
detected in the central detector and the scattered intact protons
are measured with forward detectors.

To detect the protons scattered at small angles, so-called forward
detectors are needed. The ATLAS is equipped with the Absolute
Luminosity For ATLAS (ALFA) \cite{alfa1,alfa2} and ATLAS Forward
Physics (AFP) \cite{afp1,afp2}. The CMS collaboration uses the
Precision Proton Spectrometer (PPS) as a subdetector which was born
from a collaboration between the CMS and TOTEM \cite{cms-totem}
(previously named CT-PPS). The ALFA system is made of four Roman Pot
stations located in a distance of about 240 m at both sides of the
ATLAS interaction point. The AFP detector consists of four detectors
placed symmetrically with respect to the ATLAS interaction point at
205 m (NEAR stations) and 217 m (FAR stations). The PPS detector has
four Roman Pots on each side placed symmetrically in the primary
vacuum of the LHC beam pipe, at a distance between 210 m and 220 m
from the CMS interaction point. These forward detectors are
installed as close as a few mm to the beamline to tag the intact
protons after elastic photon emission. It allows detecting the
fractional proton momentum loss in the interval
$\xi_{\min}<\xi<\xi_{\max}$. The larger value of $\xi$ can be achieved
when a detector is installed closer to the beam pipe.

Two types of examinations included by the AFP are i) exploratory
physics (anomalous couplings between $\gamma$ and $Z$ or $W$ bosons,
exclusive production, etc.); ii) standard QCD physics (double
Pomeron exchange, exclusive production in the jet channel, single
diffraction, $\gamma\gamma$ physics, etc.). PPS experiments aim at a
study of the elastic proton-proton interactions, the proton-proton
total cross-section and other diffractive processes. Moreover,
precise search can be done with the forward detectors
\cite{ttm1}-\cite{ttm3}. In such interactions involving high energy
and high luminosity, the pile-up background may be formed. This
background can be extremely reduced by using kinematics, timing
constraints, and exclusivity conditions
\cite{albrow1}-\cite{albrow3}. There are many phenomenological
papers that use photon-induced reactions for searching new physics
at the LHC \cite{inanc}-\cite{kiss1}.

We examine the process $pp \to p \gamma\gamma p \rightarrow
p'(\gamma\gamma)p'$. Emitted photons have very small virtualities,
hence they are almost-real photons and can be considered as
on-mass-shell particles. The main detector (ATLAS or CMS) registers
the final state $\gamma\gamma$, while the proton momentum loss $\xi$
is measured by the forward detector (AFP or CT-PPS). This makes it
possible to determine the invariant energy of the $\gamma\gamma$
collision, $W = 2E\sqrt{\xi_1 \xi_2}$, where $E$ is the energy of
the incoming protons. These type of collision can be studied using
equivalent photon approximation (EPA)
\cite{Terazawa:1973}-\cite{baur}. In the EPA, a photon emitted with
small angles by the protons shows the following spectrum in photon
virtuality $Q^2$ and  energy fraction $x = E_\gamma /E$,
\begin{equation}\label{photon_spectrum}
f(x, Q^{2}) = \frac{\alpha}{\pi}\frac{1}{x Q^2} \left[ ( 1 - x)
\left( 1 - \frac{Q^2_{\min}}{Q^2} \right) F_E (Q^2) +
\frac{x^2}{2}F_M(Q^2) \right] ,
\end{equation}
where
\begin{equation}\label{F_form-factors}
Q^{2}_{\min} = \frac{m^2_p \,x^2}{1 - x} \;, \quad F_{E} =
\frac{4m^{2}_{p}G^2_{E} + Q^2 G^2_{M}}{4m^{2}_p + Q^{2}} \;, \quad
F_{M} = G^{2}_{M} \;,
\end{equation}
\begin{equation}\label{G_form-factors}
G^{2}_{E} = \frac{G^2_{M}}{\mu^2_p} = \left(1 +
\frac{Q^{2}}{Q^{2}_{0}} \right)^{\!\!-4} , \quad Q^{2}_0 = 0.71 \;,
\quad \mu^{2}_{p} = 7.78 \mbox{\ GeV}^{2} \;.
\end{equation}
Here, $m_p$ is the mass of the proton, $\mu_p$ is its magnetic
moment, $F_{E}$ and $F_{M}$ are electric and magnetic form factors
of the proton. To obtain the cross section of the process
$pp\rightarrow p(\gamma\gamma)p \rightarrow p'(\gamma\gamma)p'$, the
cross section $d\sigma_{\gamma\gamma \to \gamma\gamma}$ of the
subprocess $\gamma\gamma \to \gamma\gamma$ should be integrated over
the photon spectrum,
\begin{eqnarray}\label{dif_cs}
d\sigma = \int  dW \,\frac{dL_{\gamma\gamma}}{dW}
\,d\sigma_{\gamma\gamma \rightarrow \gamma\gamma}(W) \;.
\end{eqnarray}
The effective photon luminosity in \eqref{dif_cs} is given by
\begin{equation}\label{luminosity}
\frac{dL_{\gamma\gamma}}{dW} = \frac{W}{2E^2}
\,\int_{Q^{2}_{\min}}^{Q^{2}_{max}}
{dQ^{2}_{1}}\int_{Q^{2}_{\min}}^{Q^{2}_{\max}}{dQ^{2}_{2}}
\int_{x_{\min}}^{x_{\max}} {\frac{dx}{x} \,f_{1} \!\left(
\frac{W^{2}}{4E^2x}, Q^{2}_{1} \right) f_{2}(x,Q^{2}_{2})} \;,
\end{equation}
where
\begin{equation}\label{x_min_max}
x_{\min} = \max(\xi_{\min}, W^{2}/(4E^2\xi_{\max})) \;, \quad
x_{\max} = \xi_{\max} \;.
\end{equation}
We put $Q^{2}_{\max} = 2$ $\mbox{GeV}^{2}$, since the contribution
of more than this value is very small.

\begin{figure}[htb]
\begin{center}
\includegraphics[scale=0.5]{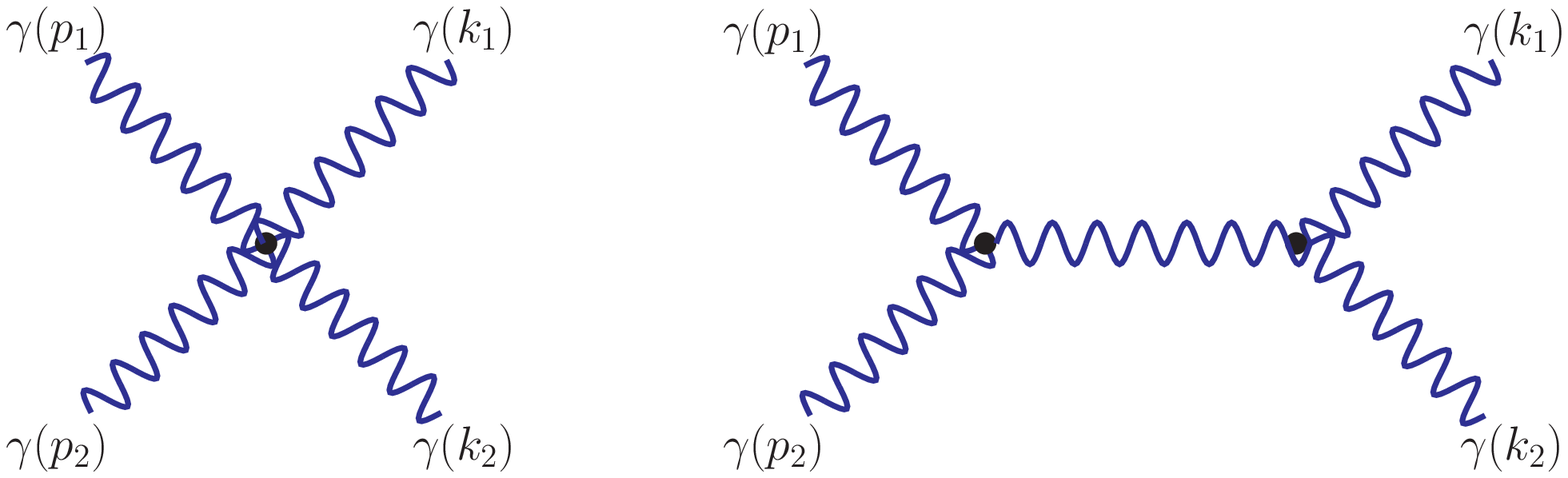} \\
\vspace{0.75cm}
\includegraphics[scale=0.5]{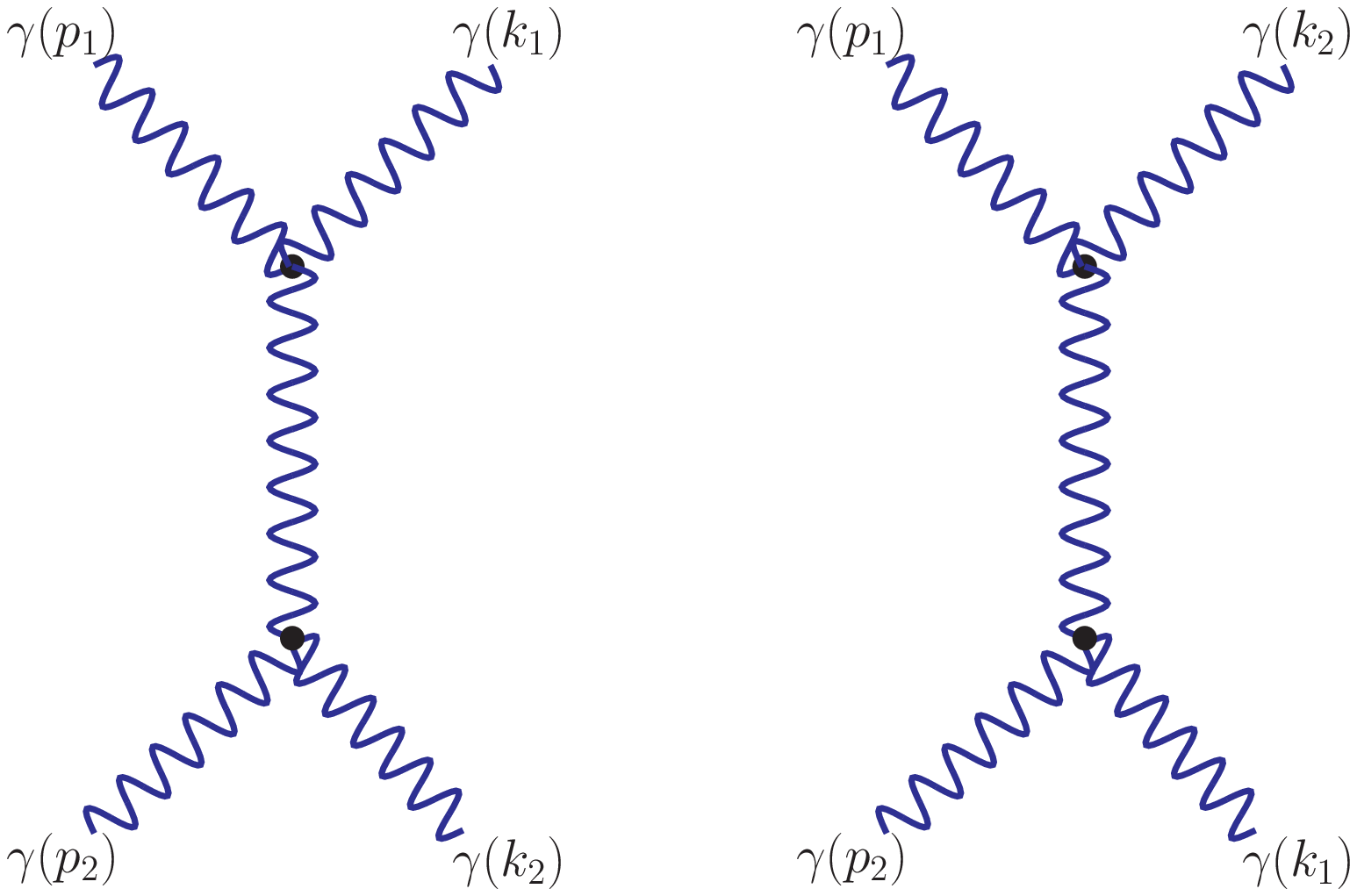}
\caption{The tree level contributions to the
$\gamma\gamma\rightarrow\gamma\gamma$ scattering in the
noncommutative QED.}
\label{fig:LBL_scatt}
\end{center}
\end{figure}

The diagrams describing the $\gamma\gamma\rightarrow\gamma\gamma$
scattering in the order $e^2$ are presented in
Fig.~\ref{fig:LBL_scatt}. The nonzero independent helicity
amplitudes of NCQED have been derived in
\cite{Hewett:2001,Latas:2021}
\begin{align}\label{++++_+--+}
M^{\mathrm{NC}}_{++++}(p_1,p_2;k_1,k_2) = -32\pi\alpha &\bigg[
\frac{s}{t} \sin \!\left( \frac{1}{2} p_1 \wedge k_1 \right) \sin
\!\left( \frac{1}{2} p_2 \wedge k_2 \right) \nonumber \\
&+ \frac{s}{u} \sin \!\left( \frac{1}{2} p_1 \wedge k_2 \right) \sin
\!\left( \frac{1}{2} p_2 \wedge k_1 \right) \bigg] ,
\nonumber \\
M^{\mathrm{NC}}_{+--+}(p_1,p_2;k_1,k_2) = - 32\pi\alpha &\bigg[
\frac{t}{s} \sin \!\left( \frac{1}{2} p_1 \wedge k_1 \right) \sin
\!\left( \frac{1}{2} p_2 \wedge k_2 \right)
\nonumber \\
&+ \frac{t^2}{su} \sin \!\left( \frac{1}{2} p_1 \wedge k_2 \right)
\sin \!\left( \frac{1}{2} p_2 \wedge k_1 \right) \bigg] ,
\end{align}
where $\alpha = e^2/(4\pi)$, and $s=(p_1 - p_2)^2$, $t=(p_1 -
k_1)^2$, $u=(p_1 - k_2)^2$ are Mandelstam variables of the
$\gamma\gamma\rightarrow\gamma\gamma$ process ($s+t+u = 0$). To
derive expressions \eqref{++++_+--+} form those presented in
\cite{Hewett:2001}, we have used the Moyal-Weyl star product Jacobi
identity in momentum space \cite{Latas:2021}
\begin{align}\label{Jacobi_identity}
&\sin \!\left( \frac{1}{2} p_1 \wedge p_2 \right) \sin \!\left(
\frac{1}{2} k_1 \wedge k_2 \right) + \sin \!\left( \frac{1}{2} p_1
\wedge k_2 \right) \sin \!\left( \frac{1}{2} p_2 \wedge k_1 \right)
\nonumber \\
&- \sin \!\left( \frac{1}{2} p_1 \wedge k_1 \right) \sin \!\left(
\frac{1}{2} p_2 \wedge k_2 \right) = 0 \;.
\end{align}

The amplitude $M^{\mathrm{NC}}_{+-+-}$ is defined by the crossing
relation
\begin{align}\label{+-+-}
&M^{\mathrm{NC}}_{+-+-} (p_1,p_2;k_1,k_2) =
M_{+--+}(p_1,p_2;k_2,k_1)
\nonumber \\
&= - 32\pi\alpha \bigg[ \frac{u^2}{st} \sin \!\left( \frac{1}{2} p_1
\wedge k_1 \right) \sin \!\left( \frac{1}{2} p_2 \wedge k_2 \right)
\nonumber \\
&\quad + \frac{u}{s} \sin \!\left( \frac{1}{2} p_1 \wedge k_2
\right) \sin \!\left( \frac{1}{2} p_2 \wedge k_1 \right) \bigg] .
\end{align}
The other nonzero NC helicity amplitudes are related to the
amplitudes \eqref{++++_+--+} and \eqref{+-+-} by the parity
relations
\begin{align}\label{amplitude_relations}
M^{\mathrm{NC}}_{----}(p_1,p_2;k_1,k_2) &=
M^{\mathrm{NC}}_{++++}(p_1,p_2;k_1,k_2) \;,
\nonumber \\
M^{\mathrm{NC}}_{-++-}(p_1,p_2;k_1,k_2) &=
M^{\mathrm{NC}}_{+--+}(p_1,p_2;k_1,k_2) \;,
\nonumber \\
M^{\mathrm{NC}}_{-+-+}(p_1,p_2;k_1,k_2) &=
M^{\mathrm{NC}}_{+-+-}(p_1,p_2;k_1,k_2) \;.
\end{align}
After simple arithmetic we find from \eqref{++++_+--+},
\eqref{+-+-}, and \eqref{amplitude_relations}
\begin{align}\label{sum_pol}
&\sum_\mathrm{pol} |M_{\mathrm{NC}}|^2 = 2 \left(
|M^{\mathrm{NC}}_{++++}|^2 + |M^{\mathrm{NC}}_{+--+}|^2 +
|M^{\mathrm{NC}}_{+-+-}|^2 \right) \nonumber \\
&= 2(32\pi\alpha)^2 \,\frac{s^4 + t^4 + u^4}{s^2} \bigg[ \frac{1}{t}
\sin \!\left( \frac{1}{2} p_1 \wedge k_1 \right) \sin \!\left(
\frac{1}{2} p_2 \wedge k_2 \right)
\nonumber \\
& \hspace{43mm} + \frac{1}{u} \sin \!\left( \frac{1}{2} p_1 \wedge
k_2 \right) \sin \!\left( \frac{1}{2} p_2 \wedge k_1 \right)
\bigg]^2 .
\end{align}
As it is shown in Appendix~A, this equation can be written as
\begin{align}\label{sum_pol_m}
&\sum_\mathrm{pol} |M_{\mathrm{NC}}|^2 = 2 \left(
|M^{\mathrm{NC}}_{++++}|^2 + |M^{\mathrm{NC}}_{+--+}|^2 +
|M^{\mathrm{NC}}_{+-+-}|^2 \right) = 2(-2)(32\pi\alpha)^2
\nonumber \\
&\times \bigg\{ \left( \frac{s}{u} + \frac{u}{s} + \frac{su}{t^2}
\right) \left[ \sin \!\left( \frac{1}{2} p_1 \wedge k_1 \right) \sin
\!\left( \frac{1}{2} p_2 \wedge k_2 \right) \right]^2
\nonumber \\
&\quad + \left( \frac{t}{s} + \frac{s}{t} + \frac{st}{u^2} \right)
\left[ \sin \!\left( \frac{1}{2} p_1 \wedge k_2 \right) \sin
\!\left( \frac{1}{2} p_2 \wedge k_1 \right) \right]^2
\nonumber \\
&\quad + \left( \frac{u}{t} + \frac{t}{u} + \frac{tu}{s^2} \right)
\left[ \sin \!\left( \frac{1}{2} p_1 \wedge p_2 \right) \sin
\!\left( \frac{1}{2} k_1 \wedge k_2 \right)\right]^2 \bigg\} ,
\end{align}
in a full agreement with eq.~(93) in \cite{Latas:2021}, after
changing momenta notations, $(p_1,p_2,k_1,k_2)\rightarrow
(k_1,k_2,k_4,k_3)$. Note that the second term in the brackets in
eq.~\eqref{sum_pol_m} is obtained from the first term if one uses
the replacements $k_1 \rightleftarrows k_2$, $t \rightleftarrows u$.
Analogously, the third term comes from the first one after the
replacements $k_1 \rightleftarrows p_2$, $t \rightleftarrows s$.

There is a one-to one correspondence between the color ordering in
QCD and the star product in pure NCQED \cite{Huang:2011}. As it was
mentioned in the end of section~2, all vertices of NCQED are similar
to gluon vertices in QCD in which the structure constants
$f^{a_ia_jc}$ are replaced by $2\sin[(l_i\wedge l_j)/2]$. A
correspondence between NCQED and QCD amplitudes can be achieved, if
one makes in \eqref{sum_pol_m} the following replacements
\cite{Latas:2021}
\begin{align}\label{NVQED_vs_QCD}
\left[ 2\sin \!\left( \frac{1}{2} l_i \wedge l_j \right) \right]^2
\left[  2\!\left( \frac{1}{2} l_k \wedge l_r \right) \right]^2
&\rightarrow \left( f^{a_i a_j c} f^{a_i a_j d} \right) \left(
f^{a_k a_r c} f^{a_k a_r d} \right)
\nonumber \\
&= 3^2 \delta^{cd}\delta^{cd} = 72 \;,
\end{align}
where $(l_i,l_j,l_k,l_r)$ is a combination of the photon momenta
$p_1,p_2,k_1,k_2$, and the sum over all color indices is assumed.
The gluon-gluon amplitude square is a sum of helicity amplitudes,
\begin{equation}\label{M_square}
|M_{gg\rightarrow gg}|^2 = \frac{1}{8^2 2^2} \sum_\mathrm{pol}
|M_{\lambda_1 \lambda_2 \lambda_3 \lambda_4}|^2 \;,
\end{equation}
and the differential cross section is given by
\begin{equation}\label{gluon_diff_cs}
\frac{d\sigma}{d\Omega}(gg\rightarrow gg) =
\frac{1}{64\pi^2}\frac{|M_{gg\rightarrow gg}|^2}{s} \;.
\end{equation}
After summation over Mandelstam variables in \eqref{sum_pol_m},
\begin{align}\label{sum_M_variables}
&\left( \frac{s}{u} + \frac{u}{s} + \frac{su}{t^2} \right) + \left(
\frac{t}{s} + \frac{s}{t} + \frac{st}{u^2} \right) + \left(
\frac{u}{t} + \frac{t}{u} + \frac{tu}{s^2} \right)
\nonumber \\
&= - \left( 3 - \frac{tu}{s^2} - \frac{su}{t^2} - \frac{ts}{u^2}
\right) ,
\end{align}
and replacement $\alpha \rightarrow \alpha_s$, we come from
\eqref{sum_pol_m}-\eqref{gluon_diff_cs} to the well-known expression
for the differential cross section of the gluon-gluon scattering
\cite{PDG}
\begin{equation}\label{gluon_cs}
\frac{d\sigma}{d\Omega}(gg\rightarrow gg) = \frac{9\alpha_s^2}{8s}
\!\left( 3 - \frac{tu}{s^2} - \frac{su}{t^2} - \frac{ts}{u^2}
\right) .
\end{equation}
Thus, our NC amplitude square and its counterpart in QCD are closely
connected.

%%%%%%%%%%%%%%%%%%%%%%%%%%%%%%%%%%%%%
% Figure 2. LBL scattering in NCQED %
%%%%%%%%%%%%%%%%%%%%%%%%%%%%%%%%%%%%%

%%%%%%%%%%%%%%%%%%%%%%%%%%%%%%
\section{Numerical analysis} %
%%%%%%%%%%%%%%%%%%%%%%%%%%%%%%

We work in the c.m.s. of the colliding protons. Let $x_1$ and $x_2$
be momentum fractions of the protons carried by the scattered
photons which can be considered to be on-shell particles. Then the
photon momenta look like
\begin{equation}\label{incoming_photon_momenta}
p_1 = x_1 E(1,0,0,1) \;, \quad p_2 = x_2 E(1,0,0,-1) \;,
\end{equation}
Thus, the $\gamma(p_1) + \gamma(p_2) \rightarrow \gamma(k_1) +
\gamma(k_2)$ collision goes in the \emph{non-center-of mass} system.
For the massless case the momenta of the outgoing photons are given
by
\begin{align}\label{outgoing_photon_momenta}
k_1 &= E_1(1, s_\theta c_\phi, s_\theta s_\phi, c_\theta) \;,
\nonumber \\
k_2 &= \left( (x_1 + x_2)E - E_1, - E_1 s_\theta c_\phi, -E_1
s_\theta s_\phi, (x_1 - x_2)E - E_1 c_\theta \right) ,
\end{align}
where
\begin{equation}\label{E1}
E_1 = \frac{2 x_1 x_2 E}{x_1 + x_2 - (x_1 - x_2)c_\theta} \;.
\end{equation}
Here $c_\theta = \cos\theta \;, \  s_\theta = \sin\theta \;, \
c_\phi = \cos\phi \;, \  s_\phi = \sin\phi$, with $\theta$ and
$\phi$ being the scattering angles of the outgoing photon with
momentum $k_1$.

Then the wage products of the photon momenta in the formulas for the
NC helicity amplitudes take the form \cite{Hewett:2001,Latas:2021}
\begin{align}\label{p_prod_k}
p_1 \wedge k_1 &= -\frac{x_1 E_1 E}{\Lambda_{\mathrm{NC}}^2} [
c_{03}(1 - c_\theta) - (c_{01} - c_{13})s_\theta c_\phi  - (c_{02} -
c_{23})s_\theta s_\phi ] \;,
\nonumber \\
p_2 \wedge k_1 &= \frac{x_2 E_1 E}{\Lambda_{\mathrm{NC}}^2} [
c_{03}(1 + c_\theta) + (c_{01} + c_{13})s_\theta c_\phi + (c_{02} +
c_{23})s_\theta s_\phi ] \;,
\nonumber \\
p_1 \wedge k_2 &= -\frac{x_1 E_1 E}{\Lambda_{\mathrm{NC}}^2} \bigg[
\frac{x_2}{x_1}(1 + c_\theta) + (c_{01} - c_{13})s_\theta c_\phi +
(c_{02} - c_{23})s_\theta s_\phi \bigg] ,
\nonumber \\
p_2 \wedge k_2 &= \frac{x_2 E_1 E}{\Lambda_{\mathrm{NC}}^2} \bigg[
\frac{x_1}{x_2}(1 - c_\theta) - (c_{01} + c_{13})s_\theta c_\phi -
(c_{02} + c_{23})s_\theta s_\phi \bigg] .
\end{align}

The Mandelstam variables of the
$\gamma\gamma\rightarrow\gamma\gamma$ process are given by
\begin{equation}\label{M_var_explicit}
s = 4E^2 x_1 x_2 \;, \quad t = -2x_1E E_1(1 - c_\theta) \;, \quad u
= -2x_2E E_1(1 + c_\theta)] \;.
\end{equation}
The differential cross section of the
$\gamma\gamma\rightarrow\gamma\gamma$ process in the frame
\eqref{incoming_photon_momenta}-\eqref{outgoing_photon_momenta} is
given by \cite{Latas:2021}
\begin{equation}\label{photon_diff_cs}
\frac{d\sigma}{d\Omega}(\gamma\gamma\rightarrow\gamma\gamma) =
\left( \frac{E_1}{4\pi s} \right)^2 \!\frac{1}{2^2}
\sum_{\mathrm{pol}} |M^{\mathrm{NC}}_{\lambda_1 \lambda_2 \lambda_3
\lambda_4} + M^{\mathrm{SM}}_{\lambda_1 \lambda_2 \lambda_3
\lambda_4}|^2 ,
\end{equation}
where $d\Omega = \sin\theta d\theta d\phi$. Note that the first
factor in the right-hand side of this equation coincides with the
conventional factor $1/(64\pi^2 s)$ only if $x_1 = x_2$.

To do our calculations, we choose the region of $0.015 < \xi < 0.15$
for both protons which is a standard acceptance in the central
detectors at the nominal accelerator and beam conditions. It is in
accordance with the intervals for $\xi$ used by the ATLAS
\cite{ATLAS_xi:2020} and CMS-TOTEM \cite{CMS-TOTEM_xi:2021}
collaborations, as well as in several papers that examine
processes with the proton tagging at the LHC
\cite{Baldenegro:2018}-\cite{Baldenegro:2020}. We also have applied
the cut on the rapidity of the outgoing photons, $|\eta|< 2.5$ in
all calculations.

The SM contribution to the $\gamma\gamma\rightarrow\gamma\gamma$
scattering is described by diagrams with charged fermion loops, $W$
boson loops, and gluon loops
\cite{Karplus:1951}-\cite{Gounaris:1999_2}. As it is shown in
\cite{Fichet:2015}, when the center of mass energy is greater than
$200$ GeV, QCD contributions are negligible, since they are very
small relative to W and fermion loops contributions. In our study,
minimum diphoton mass energy could be $2\times7000$ GeV$\times0.015$
$=$ $210$ GeV due to $\xi_{\min}=0.015$ value. Therefore, we could
safely ignore the QCD loop contributions. Explicit analytical
expressions for the SM helicity amplitudes, both for the fermion and
$W$ boson terms are too long. That is why we do not present them
here. They can be found in \cite{Inan:2020}. In such high-energy and
luminosity collisions, pile-up backgrounds can occur. With the use
of kinematics, exclusivity conditions, and timing constraints such
backgrounds can be extremely reduced \cite{albrow1, albrow2}.

As it follows from eqs.~\eqref{++++_+--+}, \eqref{+-+-},
\eqref{amplitude_relations}, and \eqref{p_prod_k}, two options,
$c_{01} = 1$, with all other $c_{\mu\nu}$ vanishing, and $c_{13} =
1$, with all other $c_{\mu\nu}$ vanishing, result in the same NC
helicity amplitudes. The same is true for the amplitudes with
$c_{02} = 1$ and $c_{23} = 1$. Given only one of the parameters
$c_{\mu\nu}$ is taken to be nonzero, the NC amplitudes are
insensitive to its sign, since they are invariant under the
$c_{\mu\nu} \rightarrow - c_{\mu\nu}$ replacement. Note also that
four-photon NC helicity amplitudes do not depend on $c_{12}$. Thus,
only three possibilities should be addressed: (i) $c_{03} = 1$, with
all other $c_{\mu\nu}$ vanishing; (ii) $c_{13} = 1$, all other NC
parameters are zero; (iii) $c_{23} = 1$, with all other $c_{\mu\nu}$
vanishing.

\begin{figure}[htb]
\begin{center}
\includegraphics[scale=0.5]{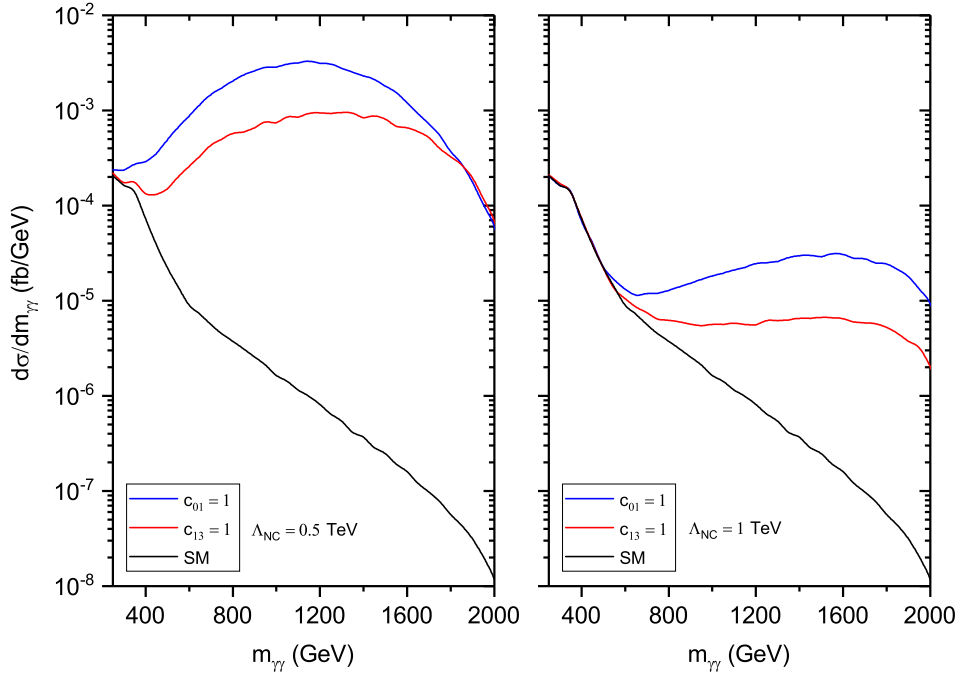} \\
\caption{The differential cross section for the $pp \rightarrow p
\gamma\gamma p \rightarrow p' \gamma\gamma p'$ scattering at the LHC
in the noncommutative QED versus invariant mass of the final
photons.}
\label{fig:MDCS}
\end{center}
\end{figure}

In Fig.~\ref{fig:MDCS} we show the results of our calculations of
the total and SM differential cross section for the diphoton
production at the 14 TeV LHC with intact protons. The cross section
is presented as a function of the invariant mass of outgoing
photons, $m_{\gamma\gamma}$, for two values of the NC scale
$\Lambda_{\mathrm{NC}}$.  We see that in the region
$m_{\gamma\gamma} > 400(600)$ GeV the NC contribution strongly
dominate the SM one for $\Lambda_{\mathrm{NC}} = 0.5(1.0)$ TeV. For
both values of the NC scale, the time-space NC (blue curves in
Fig.~\ref{fig:MDCS}) is larger that the space-space NC (red curves).
Only for $\Lambda_{\mathrm{NC}} = 0.5$ TeV and $m_{\gamma\gamma} >
1800$ GeV the time-space and space-space curves merge. From
eqs.~\eqref{++++_+--+}, \eqref{+-+-}, and \eqref{p_prod_k} one makes
sure that after integrations over angular variables, the NC
amplitudes should give the same contribution to the differential
cross section both for $c_{13} = 1$ and for $c_{23} = 1$. Our
numerical calculations confirm this statement. That is why, we do
not present curves for case (iii) in Fig.~\ref{fig:MDCS}.
%
%%%%%%%%%%%%%%%%%%%%%%%%%%%%%%%%%%%%%5%%%
% Figure 3. Differential cross sections %
%%%%%%%%%%%%%%%%%%%%%%%%%%%%%%%%%%%%%%%%%

\begin{figure}[htb]
\begin{center}
\includegraphics[scale=0.5]{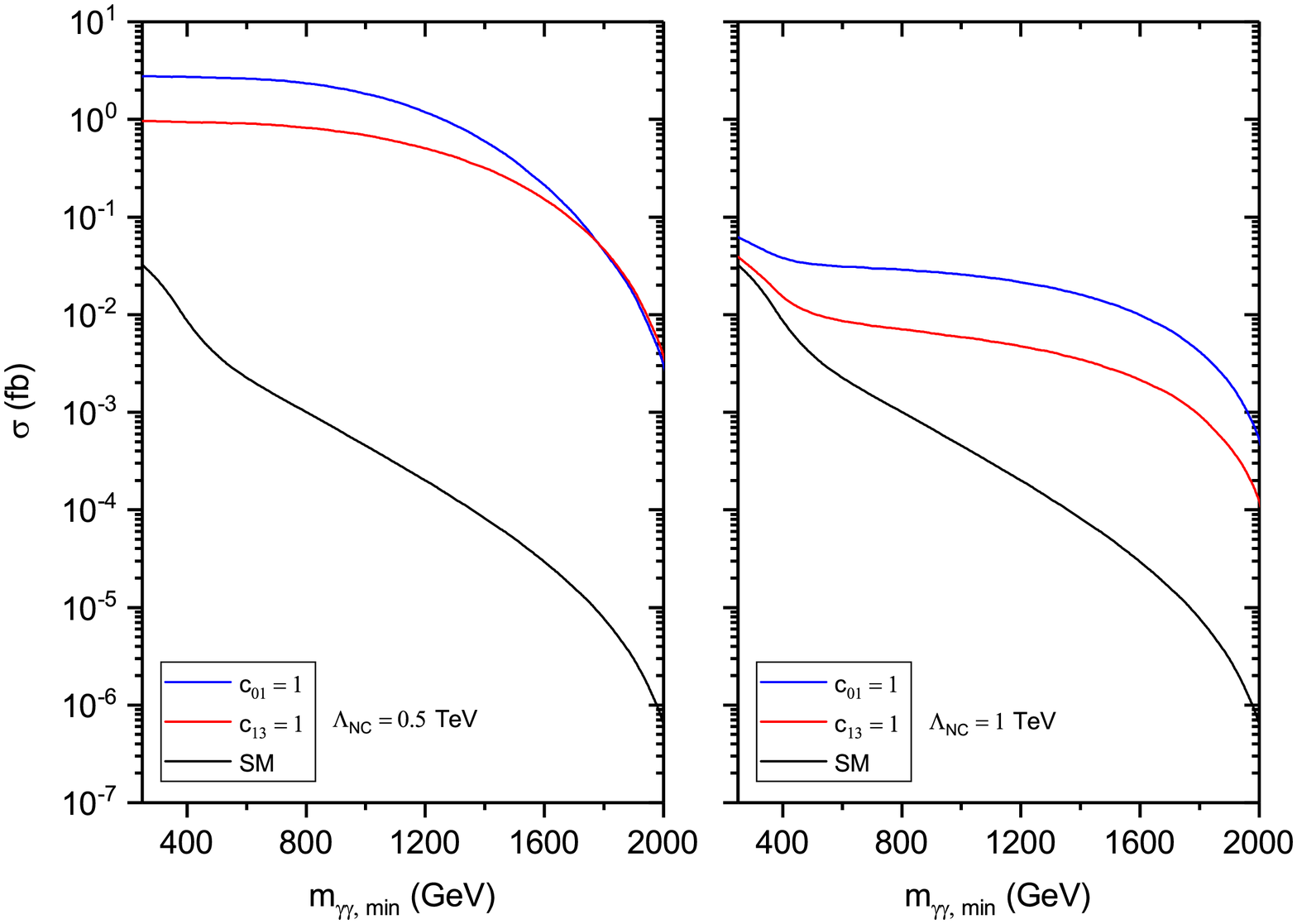} \\
\caption{The cross section $\sigma(m_{\gamma\gamma} \!>
\!m_{\gamma\gamma,\,\min})$ for the $pp \rightarrow p \gamma\gamma p
\rightarrow p' \gamma\gamma p'$ scattering at the LHC in the
noncommutative QED versus minimal invariant mass of the diphoton
system.}
\label{fig:MCUTCS}
\end{center}
\end{figure}

In Fig.~\ref{fig:MCUTCS} the total and SM cross sections
$\sigma(m_{\gamma\gamma} \!> \!m_{\gamma\gamma,\,\min})$ versus
$m_{\gamma\gamma,\,\min}$, the minimal invariant mass of the
outgoing photons, is presented. As one can see, for
$\Lambda_{\mathrm{NC}} = 0.5$ TeV this cross section is
approximately two order of magnitude larger than the SM cross
section in all mass region. For $\Lambda_{\mathrm{NC}} = 1$ TeV the
deviation of the cross section from the SM one is also very large
and becomes more and more prominent as $m_{\gamma\gamma,\,\min}$
grows. Thus, the bigger is the value of $m_{\gamma\gamma,\,\min}$,
the larger is the difference between the new physics and SM. Note
that the time-space NC exceeds the space-space NC for all
$m_{\gamma\gamma,\,\min}$. Only for $\Lambda_{\mathrm{NC}} = 0.5$
TeV it becomes comparable with the space-space NC in the large
$m_{\gamma\gamma,\,\min}$ region.
%
%%%%%%%%%%%%%%%%%%%%%%%%%%%%%%%%%%
% Figure 4. Total cross sections %
%%%%%%%%%%%%%%%%%%%%%%%%%%%%%%%%%%

\begin{figure}[htb]
\begin{center}
\includegraphics[scale=0.5]{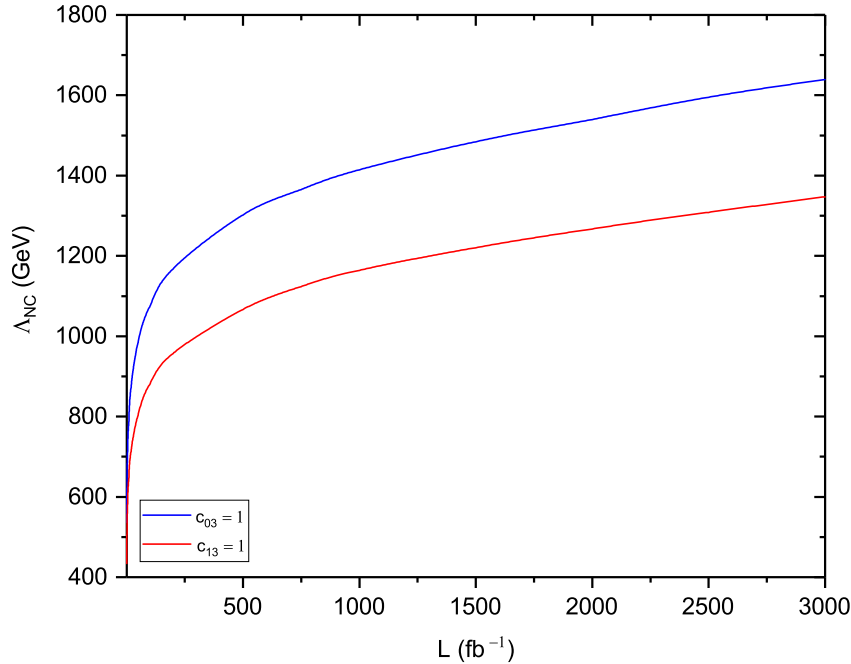} \\
\caption{95\% C.L. bounds on the noncommutative scale
$\Lambda_{\mathrm{NC}}$ coming from the $pp \rightarrow p
\gamma\gamma p \rightarrow p' \gamma\gamma p'$ scattering as
functions of integrated luminosity of proton-proton collision at 14
TeV. The curves are obtained with the use of the cut on diphoton
invariant mass, $m_{\gamma\gamma} > 1000$ GeV.}
\label{fig:NCBOUNDS}
\end{center}
\end{figure}

Knowing cross sections and SM backgrounds, we have calculated upper
bounds on the NC scale $\Lambda_{\mathrm{NC}}$ which can be obtained
from the $pp \rightarrow p \gamma\gamma p \rightarrow p'
\gamma\gamma p'$ scattering at the LHC.
To obtain the exclusion
region, we applied the following equation for the statistical
significance (SS) \cite{SS}
\begin{equation}\label{SS_def}
SS = \sqrt{2[(S - B\,\ln(1 + S/B)]} \;,
\end{equation}
where $S$ is the number of signal events and $B$ is the number of
background events. We define the regions $SS \leqslant 1.645$ as the
regions that can be excluded at the 95\% C.L. To reduce the SM
background, we used the cut $m_{\gamma\gamma} > 1000$ GeV . The
results are shown in Fig.~\ref{fig:NCBOUNDS}. We see that two-photon
process at the LHC can probe the NC scale $\Lambda_{\mathrm{NC}}$ up
to 1.64(1.35) TeV for time-space (space-space) NC parameters for the
integrated luminosity $L = 3000$ fb$^{-1}$. If $L = 1000$ fb$^{-1}$,
the upper bounds are approximately 1.40 TeV and 1.15 TeV, for
time-space and space-space NC parameters, respectively.
%%%%%%%%%%%%%%%%%%%%%%%%%%%%%%%%
% Figure 5. Bounds on NC scale %
%%%%%%%%%%%%%%%%%%%%%%%%%%%%%%%%

%%%%%%%%%%%%%%%%%%%%%%%
\section{Conclusions} %
%%%%%%%%%%%%%%%%%%%%%%%
We have examined the light-by-light scattering in the noncommutative
(NC) quantum electrodynamics (NCQED) in the $pp \rightarrow p
\gamma\gamma p \rightarrow p' \gamma\gamma p'$ collision at the 14
TeV LHC. The NC geometry of space-time appears within a
framework of string theory. NCQED is based on the $U(1)$ group, but
it exhibits the non-Abelian nature in having both 3-photon and
4-photon vertices in the lagrangian. That is why the
$\gamma\gamma\rightarrow\gamma\gamma$ scattering must be sensitive
to the NC contributions at the tree level, see
Fig.~\ref{fig:LBL_scatt}.

We have calculated the differential cross sections as functions of
the invariant mass of the outgoing photons (Fig.~\ref{fig:MDCS}), as
well as the cross sections as functions of the minimum invariant
mass of the outgoing photons (Fig.~\ref{fig:MCUTCS}). Both
time-space and space-space NC parameters have been considered. The
SM background is defined by the contributions from the $W$ and
charged fermion loops. It allowed us to estimate the 95\% C.L.
bounds for the NC scale $\Lambda_{\mathrm{NC}}$. We have shown that
the scales up to $\Lambda_{\mathrm{NC}} = 1.64(1.35)$ TeV can be
probed at the LHC, for the time-space (space-space) NC parameters,
see Fig.~\ref{fig:NCBOUNDS}. Our bounds are stronger than the limits
that can be obtained in the 4-photon scattering at high energy
linear colliders \cite{Hewett:2001}. Note that
$\gamma\gamma\rightarrow\gamma\gamma$ is the more appropriate
channel to probe $\Lambda_{\mathrm{NC}}$ also in Pb-Pb collisions at
the LHC \cite{Latas:2021}.

%%%%%%%%%%%%%%%%%%%%%%%%%%%%
\section*{Acknowledgments} %
%%%%%%%%%%%%%%%%%%%%%%%%%%%%

We would like to thank Josip Trampeti\'{c} for discussions and
sending us the file with the outcome of calculations of
noncommutative helicity amplitudes.

%%%%%%%%%%%%%%%%%%%%%%%%%%%%%%%%%%%%%%%%%%%%%%%%%%%%%%%%%%%%%%%%%%%%%

%%%%%%%%%%%%%%
% Appendix A %
%%%%%%%%%%%%%%

\setcounter{equation}{0}
\renewcommand{\theequation}{A.\arabic{equation}}

\section*{Appendix A}
\label{app:A}

The right-hand side of eq.~\eqref{sum_pol}, with the factor
$2(32\pi\alpha)^2$ omitted, looks like
\begin{align}\label{I_1}
I = \frac{s^4 + t^4 + u^4}{s^2} &\bigg[ \frac{1}{t} \sin \!\left(
\frac{1}{2} p_1 \wedge k_1 \right) \sin \!\left( \frac{1}{2} p_2
\wedge k_2 \right)
\nonumber \\
&+ \frac{1}{u} \sin \!\left( \frac{1}{2} p_1 \wedge k_2 \right) \sin
\!\left( \frac{1}{2} p_2 \wedge k_1 \right) \bigg]^2 .
\end{align}
From the Jacobi identity \eqref{Jacobi_identity} it follows that
\begin{align}\label{J_prod}
&2 \left[ \sin \!\left( \frac{1}{2} p_1 \wedge k_1 \right) \sin
\!\left( \frac{1}{2} p_2 \wedge k_2 \right) \right] \!\times
\!\left[ \sin \!\left( \frac{1}{2} p_1 \wedge k_2 \right) \sin
\!\left( \frac{1}{2} p_2 \wedge k_1 \right) \right]
\nonumber \\
&= \left[ \sin \!\left( \frac{1}{2} p_1 \wedge k_1 \right) \sin
\!\left( \frac{1}{2} p_2 \wedge k_2 \right) \right]^2 + \left[ \sin
\!\left( \frac{1}{2} p_1 \wedge k_2 \right) \sin \!\left(
\frac{1}{2} p_2 \wedge k_1 \right) \right]^2
\nonumber \\
&- \left[ \sin \!\left( \frac{1}{2} p_1 \wedge p_2 \right) \sin
\!\left( \frac{1}{2} k_1 \wedge k_2 \right)\right]^2 .
\end{align}
Then we obtain from \eqref{I_1}
\begin{align}\label{I_2}
I = -\frac{s^4 + t^4 + u^4}{s tu} &\bigg\{ \frac{1}{t} \left[ \sin
\!\left( \frac{1}{2} p_1 \wedge k_1 \right) \sin \!\left(
\frac{1}{2} p_2 \wedge k_2 \right) \right]^2
\nonumber \\
&+ \frac{1}{u} \left[ \sin \!\left( \frac{1}{2} p_1 \wedge k_2
\right) \sin \!\left( \frac{1}{2} p_2 \wedge k_1 \right) \right]^2
\nonumber \\
&+ \frac{1}{s} \left[ \sin \!\left( \frac{1}{2} p_1 \wedge p_2
\right) \sin \left( \frac{1}{2} k_1 \wedge k_2 \right)\right]^2
\bigg\} .
\end{align}
Using relation $s^4 + t^4 + u^4 = 2(s^2t^2 + t^2u^2 + u^2s^2)$, we
get the equation
\begin{align}\label{I_3}
I = (-2)&\bigg\{ \left( \frac{s}{u} + \frac{u}{s} + \frac{su}{t^2}
\right) \left[ \sin \!\left( \frac{1}{2} p_1 \wedge k_1 \right) \sin
\!\left( \frac{1}{2} p_2 \wedge k_2 \right) \right]^2
\nonumber \\
&+ \left( \frac{t}{s} + \frac{s}{t} + \frac{st}{u^2} \right) \left[
\sin \!\left( \frac{1}{2} p_1 \wedge k_2 \right) \sin \!\left(
\frac{1}{2} p_2 \wedge k_1 \right) \right]^2
\nonumber \\
&+ \left( \frac{u}{t} + \frac{t}{u} + \frac{tu}{s^2} \right) \left[
\sin \!\left( \frac{1}{2} p_1 \wedge p_2 \right) \sin \!\left(
\frac{1}{2} k_1 \wedge k_2 \right)\right]^2 \bigg\} ,
\end{align}
that results in formula \eqref{sum_pol_m}.

%%%%%%%%%%%%%%%%%%%%%%%%%%%%%%%%%%%%%%%%%%%%%%%%%%%%%%%%%%%%%%%%%%%%%

%%%%%%%%%%%%%%
% References %
%%%%%%%%%%%%%%

%%%%%%%%%%%%%%%%%%%%%%%%%%%%%%%%%%%%%%%%%%%%%%%%%%%%%%%%%%%%%%%%%%%%%

%%%%%%%%%%%%%%%
\end{document}